\documentclass[baaa]{baaa}

\usepackage[pdftex]{hyperref}
\usepackage{subfigure}
\usepackage{natbib}
\usepackage{helvet,soul}
\usepackage[font=small]{caption}

\begin{document}


\journalvol{61A}
\journalyear{2019}
\journaleditors{R. Gamen, N. Padilla, C. Parisi, F. Iglesias \& M. Sgr\'o}


\contriblanguage{1}


\contribtype{2}

\thematicarea{6}

\title{Synthetic observations of H$_2$D$^+$ towards high-mass starless cores}
\subtitle{}


\titlerunning{Synthetic observations of high-mass starless cores}


\author{J. Zamponi\inst{1}, D.~R.~G. Schleicher\inst{1}, S. Bovino\inst{1}, A. Giannetti\inst{2}, G. Sabatini\inst{1,2,3} \& S. Ferrada-Chamorro\inst{1}}
\authorrunning{Zamponi et al.}


\contact{jzamponi@udec.cl}

\institute{
            Departamento de Astronom\'ia, Universidad de Concepci\'on,
            Esteban Iturra s/n Barrio universitario, Casillo 160-C, Concepci\'on, Chile
            \and
            INAF-Istituto di Radioastronomia - Italian ARC, Via P. Gobetti, 101, I-40129 Bologna, Italy
            \and 
            Dipartimento di Fisica e Astronomia, Università degli Studi di Bologna, via Gobetti 93/2, I-40129 Bologna, Italy
}


\resumen{
    Las estrellas masivas jóvenes residen comúnmente en cúmulos moleculares densos y masivos y son conocidas por estar altamente oscurecidas y distantes. Durante su proceso de formación, la deuteración es considerada como un buen indicador de la etapa de formación en la que se encuentra un objeto. Observaciones adecuadas de regiones con emisión de deuterio son cruciales, aunque difíciles de realizar. En este trabajo se hizo un survey para detectar la transición o-H$_2$D$^+(1_{10}$-$1_{11})$ en moléculas deuteradas, utilizando una fuente sintética e intentando declarar cuán diferente es la información obtenida por un interferómetro o un telescopio de disco simple. Con objeto de analizar la detectabilidad de esta transición, procesamos la simulación magneto-hidrodinámica de una nube colapsante haciendo uso del código de transferencia radiativa POLARIS. 
    Utilizando los mapas de intensidad resultantes, realizamos observaciones sintéticas de tipo interferométricas (ALMA) y de disco simple (APEX) de una nube en varios estados evolutivos, siempre usando modelos realistas. Finalmente, derivamos densidades de columna para comparar nuestras simulaciones con observaciones anteriormente realizadas. 
    Las densidades de columna obtenidas para el o-H$_2$D$^+$ concuerdan con valores reportados en la literatura, en el rango de 10$^{10-11}$cm$^{\!-2}$ y 10$^{12-13}$cm$^{\!-2}$ en mediciones interferométricas y de disco simple.
}

\abstract{
    Young massive stars are usually found embedded in dense and massive molecular clumps and are known for being highly obscured and distant. During their formation process, deuteration is regarded as a potentially good indicator of the formation stage. Therefore, proper observations of such deuterated molecules are crucial, but still, hard to perform. In this work, we test the observability of the transition o-H$_2$D$^+(1_{10}$-$1_{11})$, using a synthetic source, to understand how the physical characteristics are reflected in observations through interferometers and single-dish telescopes. In order to perform such  tests, we post-processed a magneto-hydrodynamic simulation of a collapsing magnetized core using the radiative transfer code POLARIS. 
    Using the resulting intensity distributions as input, we performed single-dish (APEX) and interferometric (ALMA) synthetic observations at different evolutionary times, always mimicking realistic configurations. Finally, column densities were derived to compare our simulations with real observations previously performed. Our derivations for o-H$_2$D$^+$ are in agreement with values reported in the literature, in the range of 10$^{\!10-11}$cm$^{\!-2}$ and 10$^{\!12-13}$cm$^{\!-2}$ for single-dish and interferometric measurements, respectively.
}


\maketitle

\section{Introduction}
\label{sec:introduction}
Massive star formation takes place in the the densest part of molecular clouds, which are characterized by low gas temperatures ($T_{\rm g}<$20K), high column densities ($N_{\rm g}\!\sim\!10^{23-25}$~cm$^{\!-2}$) and a large degree of CO-depletion. In the earlier stages, due to the latter process, further chemical reactions become much more efficient, such as the formation of deuterated molecules. This process is called \textit{deuterium fractionation} ($D_{\rm frac}$) and explains the increased ratio of a deuterated isotopologue column density and its hydrogenated version. 
The reactions that lead to deuteration are the following:

\begin{eqnarray*}
{\rm H}_3^+ +{\rm HD}&\rightleftharpoons&{\rm H}_2{\rm D}^+ +{\rm H}_2 +230~{\rm K},\\
{\rm H}_2{\rm D}^+ +{\rm HD}&\rightleftharpoons&{\rm D}_2{\rm H}^+ +{\rm H}_2 +180~{\rm K},\\
{\rm D}_2{\rm H}^+ +{\rm HD}&\rightleftharpoons&{\rm D}_3^+ +{\rm H}_2 +230~{\rm K},
\end{eqnarray*}
where H$_2$D$^+$ and D$_2$H$^+$ are the first two deuterated ions created. H$_2$D$^+$, in particular, has been suggested to be a reliable chemical clock for star forming regions, due to its sensitivity to environmental conditions (e.g., \citet{Caselli03, Brunken14}).


Observations of several high-mass clumps have shown that deuteration increases over time as the collapse of the molecular clump proceeds, reaching a maximum point right before the formation of a protostellar object \citep{Fontani11}. At this stage, the radiation from the stellar object will lead to an increase of temperatures, with subsequent evaporation of CO, and an eventual decrease of the deuteration fraction.




In this work we traced the abundance of o-H$_2$D$^+$, by performing synthetic observations towards simulated massive starless cores, and compare the results with available observations.

\begin{figure}
      \centering
      \includegraphics[width=8.5cm,trim={2.5cm 4cm 2.5cm 7cm},clip]{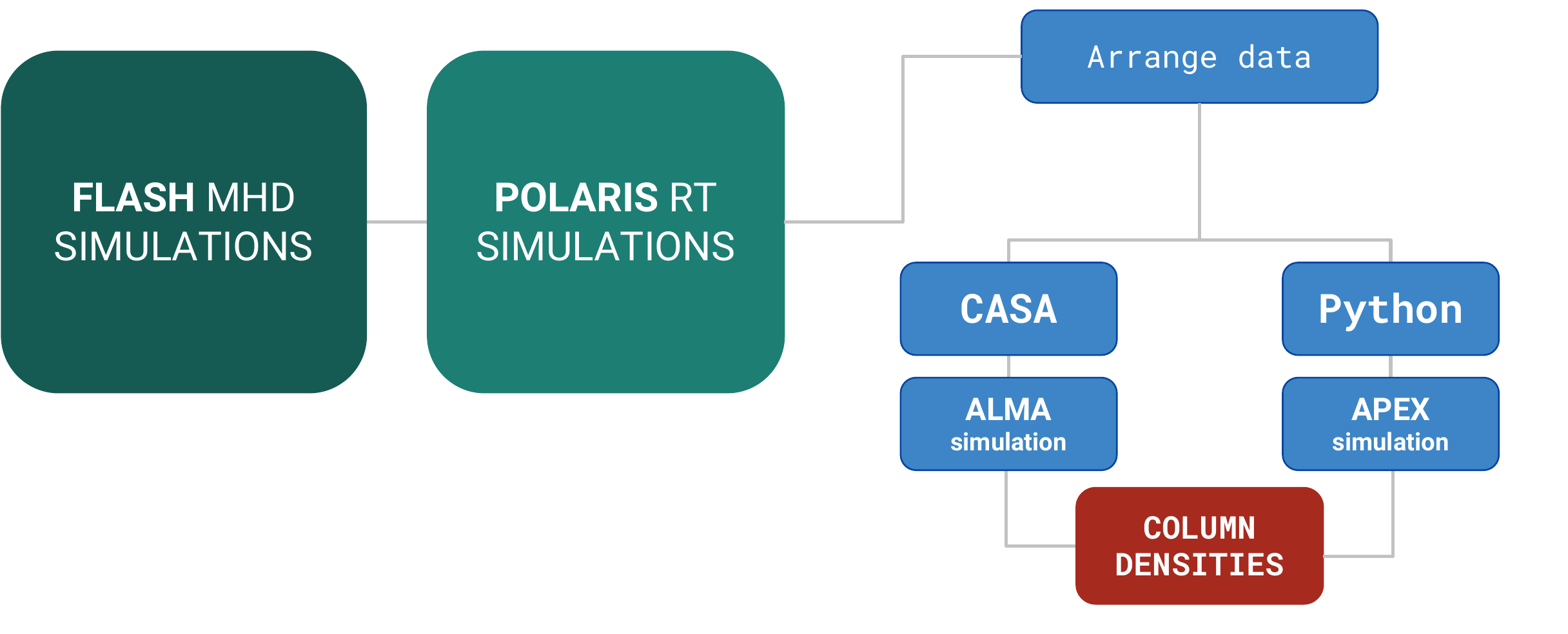}
      \caption{Flowchart of the pipeline workflow.}
      \label{fig:pipeline_flowchart}
  \end{figure}

Fig.~\ref{fig:pipeline_flowchart} shows a schematic view of the workflow followed here, from the acquisition of the source, to the final derived observatory-dependent column densities.

\section{Synthetic source}
 As synthetic source we used the simulated collapsing molecular cores from \citet{Kortgen17}. 

\begin{figure*}
    \centering
    \begin{tabular}{c c c c}
        \hspace{1cm}MHD MODEL\hspace{0.6cm} & RADIATIVE TRANSFER\hspace{0.85cm} & ALMA~(1")\hspace{2.05cm} & APEX~(16.8")\hspace{1.2cm}
    \end{tabular}\\
    \hspace{-0.7cm}
    \includegraphics[width=3.5cm,trim={2.9cm 2.3cm 4.5cm 2cm},clip]{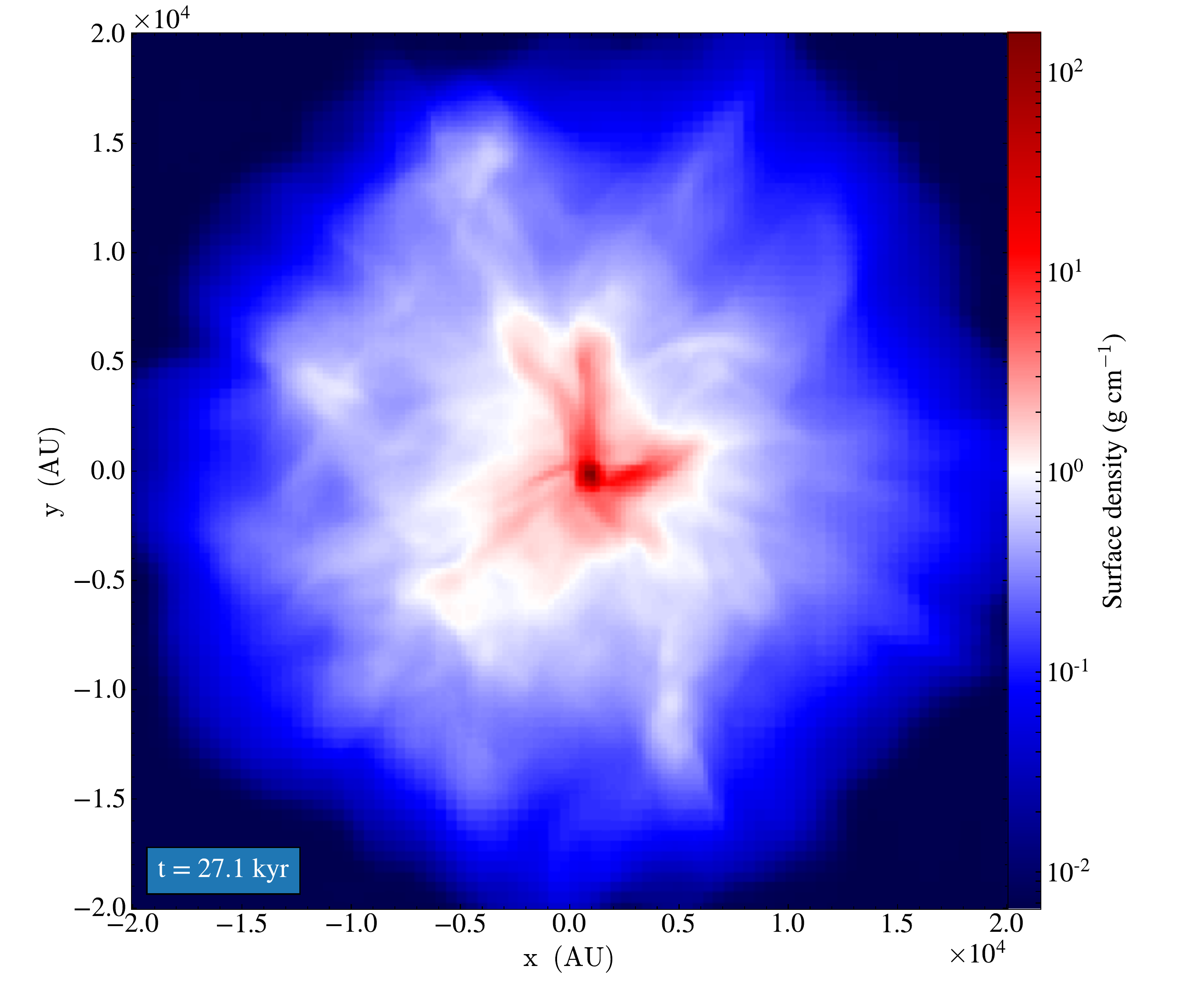}
    \hspace{.4cm}
    \includegraphics[width=3.45cm,trim={2cm 2cm 3.5cm 1.2cm},clip]{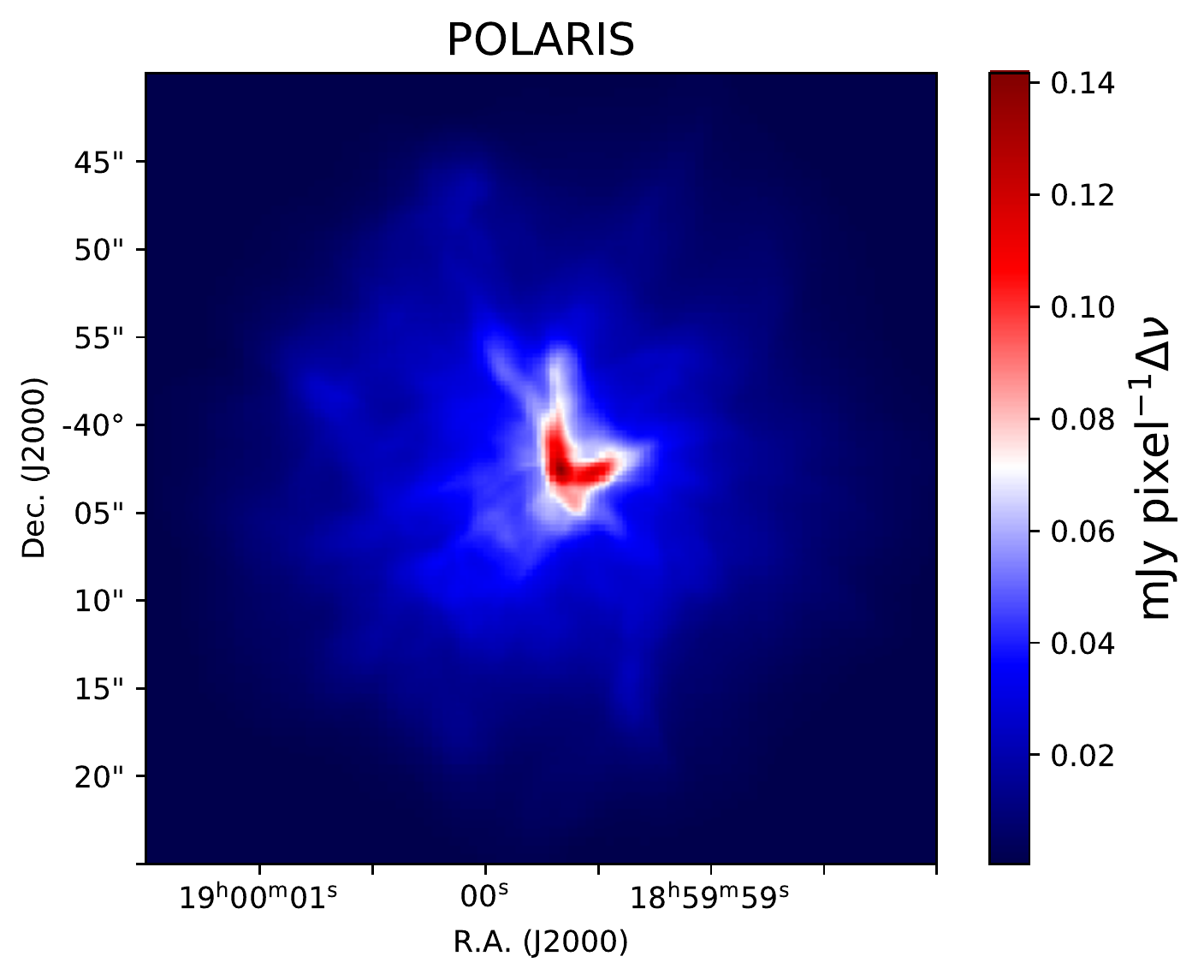}
    \hspace{.4cm}
    \includegraphics[width=3.32cm,trim={2cm 2cm 3.5cm 1.2cm},clip]{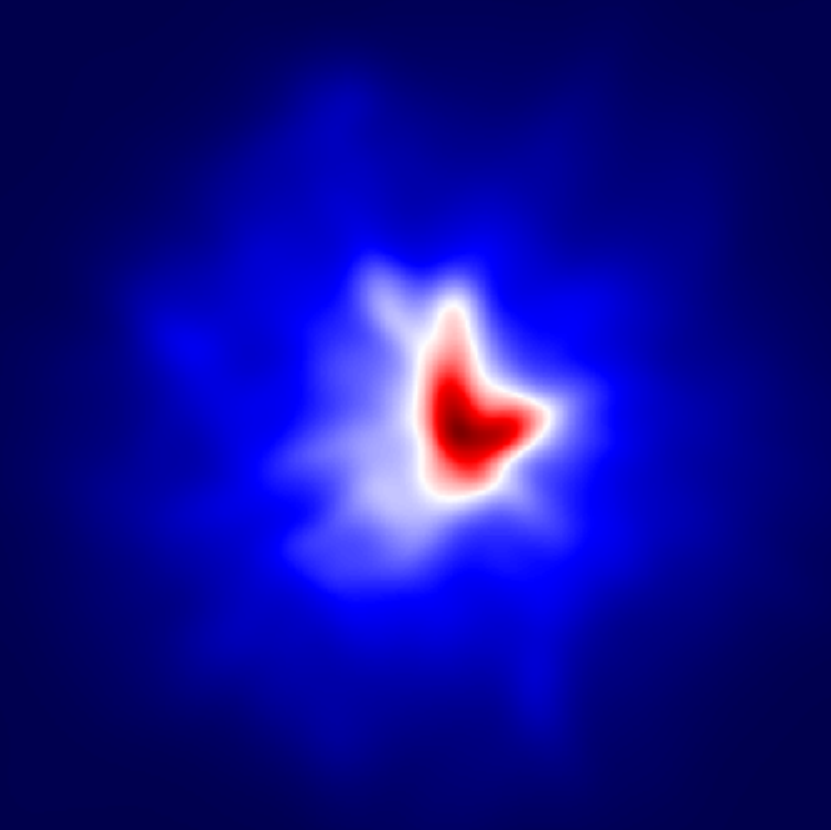}
    \hspace{.4cm}
    \includegraphics[width=4.02cm,trim={2cm 2cm 2cm 1.2cm},clip]{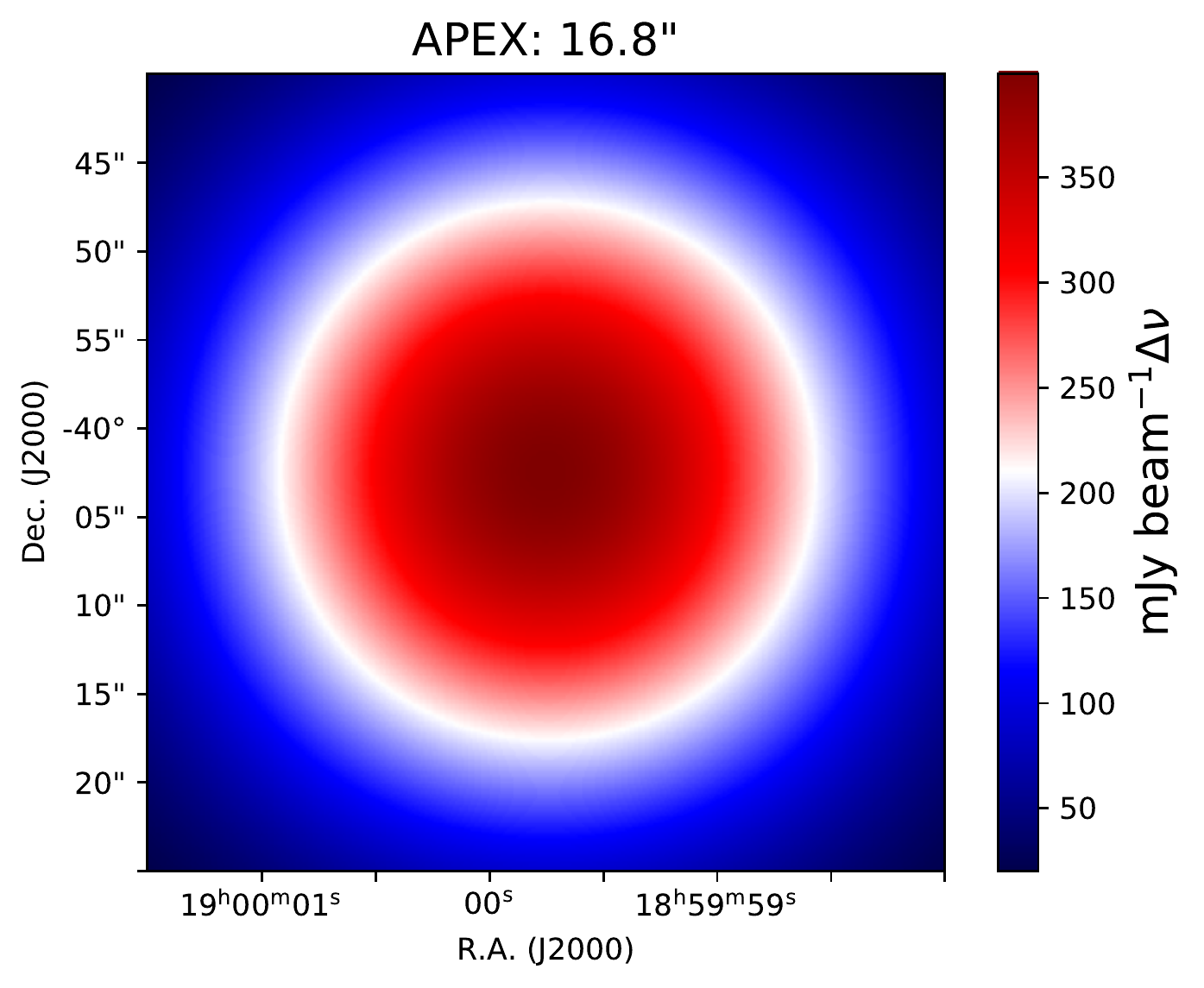}
    \hspace{-.5cm}

    \caption{\textit{Left to right:} Gas surface density distribution (10$^{-2}$-10$^2$~g~cm$^{-1}$) of the clump used as synthetic source at 27.1~kyr; o-H$_2$D$^+$ emission after the radiative transfer simulation (0.02-0.14~mJy~pixel$^{-1}$); ALMA synthetic observation (5-45~mJy~beam$^{-1}$) of the clump at 1" resolution and APEX synthetic observation (50-350~mJy~beam$^{-1}$) at 16.8" resolution. All panels share field of view of 40kAU$\sim$28".}
    \label{fig:mhd_snapshots}
\end{figure*}

 These cores are turbulent and magnetized ($B\!\sim\! 70\mu$G), 60 M$_\odot$ in mass, and 0.1 pc in radius. The main parameters of the cores are listed in Table~\ref{tab:mhd_params}. A snapshot of their gas surface density distribution after 32.1 kyr of evolution is shown in the left panel of Fig.~\ref{fig:mhd_snapshots}. 
 \begin{table}
    \centering
    \caption{Initial parameters of the core selected. The collapse is isothermal, at T=15~K.}
    \label{tab:mhd_params}
    \begin{tabular}{lccccc}
    \hline
    Run & Radius & Mass & $t_{\rm ff}$ & $\alpha_{\rm vir}$ & $\mathbf{\mathcal{M}}$ \\
      & (pc) & (M$_{\odot}$) & (kyr) &  & \\ 
    \hline
    \hline
    Hmu10M2 & 0.1  & 60 & 67 & 0.48 & 2  \\ 
    \hline
    \end{tabular}
\end{table}
 
\section{Radiative Transfer}
We employed the \textit{POLArized RadIation Simulator} (POLARIS) radiative transfer code \citep{Reissl16} to simulate the resulting intensity distributions of several quantities included in the cloud simulation, such as, temperature, dust, gas and magnetic field distribution. POLARIS makes use of a Monte Carlo approach to trace the path of light rays before reaching the synthetic detector. This method significantly reduces the numerical noise as well as the runtime  compared to previous methods \citep{Haworth08}. To map the deuterated molecule distributions, the Line Radiative Transfer (LRT) simulation mode of POLARIS was used for the o-H$_2$D$^+$ transition $(1_{10}$-$1_{11})$, assuming Local Thermodynamical Equilibrium (LTE) conditions. We analyzed 11 evolutionary stages of the core up to 1~$t_{\rm ff}$, placing it at a distance of 1.4~kpc, based on the distance to a similar source already observed \citep{Pillai12}. The resulting intensity distribution of a POLARIS simulation is shown in the second panel of Fig.~\ref{fig:mhd_snapshots}.
\section{Synthetic observations}
To make observations as realistic as possible, we post-processed the ideal synthetic maps from POLARIS, in order to take into account instrument-related effects. 

\subsection{Single-dish observations}
For the single-dish case, new tasks were written to solve each step involved in the creation of realistic cubes. The Simulations presented here are referred to as APEX-like observations, as we do not attempt to make statements of the \textit{Atacama Pathfinder Experiment} (APEX) real throughput and performance. We developed a Python module that allows to perform common tasks on data cubes and single images, such as the convolution with a Point Spread Function (PSF), conversion between intensities, fluxes and brightness temperatures and the addition of noise. We first convolved the ideal images with the beam of the telescope, represented by the PSF, here assumed to be Gaussian. The beam resolution was 16.8'' at 372.4~GHz ($\Delta v$=$0.03~\rm km\,s^{-1}$). Then we added normally-distributed noise to the images. In order to increase the peak signal-to-noise ratio (SNR), the spectra were binned up to 0.5 $\rm km\,s^{-1}$, resulting in a noise level ($T_{\rm rms}$) of 0.02~K. Such a setup returns detections at a 5$\sigma$ confidence level, for an integration of 6~hr. The values were computed using the \textit{APEX Observing Time Calculator}. See the rightmost panel in Fig.~\ref{fig:mhd_snapshots} for a 16.8" single-pointing synthetic observation.

\subsection{Interferometric observations}
For the interferometric observations, we used the \textit{Common Astronomy Software Applications} (CASA). CASA has been designed for the \textit{Atacama Large Milimiter-submiliter Array} (ALMA) and the \textit{Very Large Array} (VLA) data analysis, and provides also a simulation mode. As with the single-dish data, we used the POLARIS outcome as the input model and then generated the measurement set by calling the \texttt{simobserve} task. \texttt{simobserve} simulates an actual observation by creating a visibility set. We used a synthesized beam of 1" for o-H$_2$D$^+$, reached by the Cycle 6 C43-1 array configuration in band 7. In this case, we set the ALMA spectral resolution one order of magnitude wider than APEX ($0.3~\rm km\,s^{-1}$), in order to improve the sensitivity. The noise level was set at 6~mJy for an integration time of 40~min, obtained by the \textit{ALMA Sensitivity Calculator}. We imaged the data by deconvolving the visibilities using the CLEAN algorithm. The cleaning step was performed by the \texttt{simanalyze} task. The outcome of an ALMA simulation is shown in the third panel of Fig.~\ref{fig:mhd_snapshots}.

\section{Results \& Conclusions}
\subsection{Column densities}

Column densities of o-H$_2$D$^+$ were obtained as in \citet{Vastel06}, via
\begin{equation*}
    \label{eq:column_density}
    N(X) = \frac{8\pi\nu^3}{c^3}\frac{Q(T_{\rm ex})}{g_{u}A_{ul}}\frac{e^{E_{u}/T_{\rm ex}}}{e^{h\nu/kT_{\rm ex}}-1}\int\tau dv,
\end{equation*}
where $u$ and $l$ refer to the upper and lower energy level of each transition, respectively. All of the parameters required here are retrieved from the Leiden Atomic and Molecular Database  (LAMDA) \citep{Schoier05}, and summarized in Table~\ref{tab:line_params}. The partition function ($Q$) was recreated by a cubic spline interpolation of the available values in the Cologne Database for Molecular Spectroscopy\footnote{\url{http://www.astro.uni-koeln.de/cdms}} (CDMS). Under the LTE assumption, the excitation temperature ($T_{\rm ex}$) is assumed to match $T_{\rm gas}$.

\begin{table}[]
    \caption{LAMDA parameters for the o-H$_2$D$^+(1_{10}$-$1_{11})$ transition. The statistical weight $g_{ul}$ is 9 and $T_{\rm ex}$ 15~K.}
    \label{tab:line_params}
    \centering
    \begin{tabular}{lcccc}
    \hline
    Transition & \!\!\!\!\!$\nu$~(GHz) & \!\!\!\!\!$A_{ul}$ (s\!$^{\rm-1}$) & \!\!\!\!$E_{ul}$~(K) & \!\!\!Q($T_{\rm ex}$) \\
    \hline
    \hline
    o-H$_2$D$^+(1_{10}$-$1_{11})$ & \!\!\!\!\!372.421 & \!\!\!\!\!1.08$\cdot$10$^{-4}$ & \!\!\!\!17.87 & \!\!\!1.122 \\
    \hline
    \end{tabular}
\end{table}

\begin{figure}
    \centering
    \includegraphics[width=8.5cm,trim={0.0cm 0.0cm 0.0cm 0.0cm},clip]{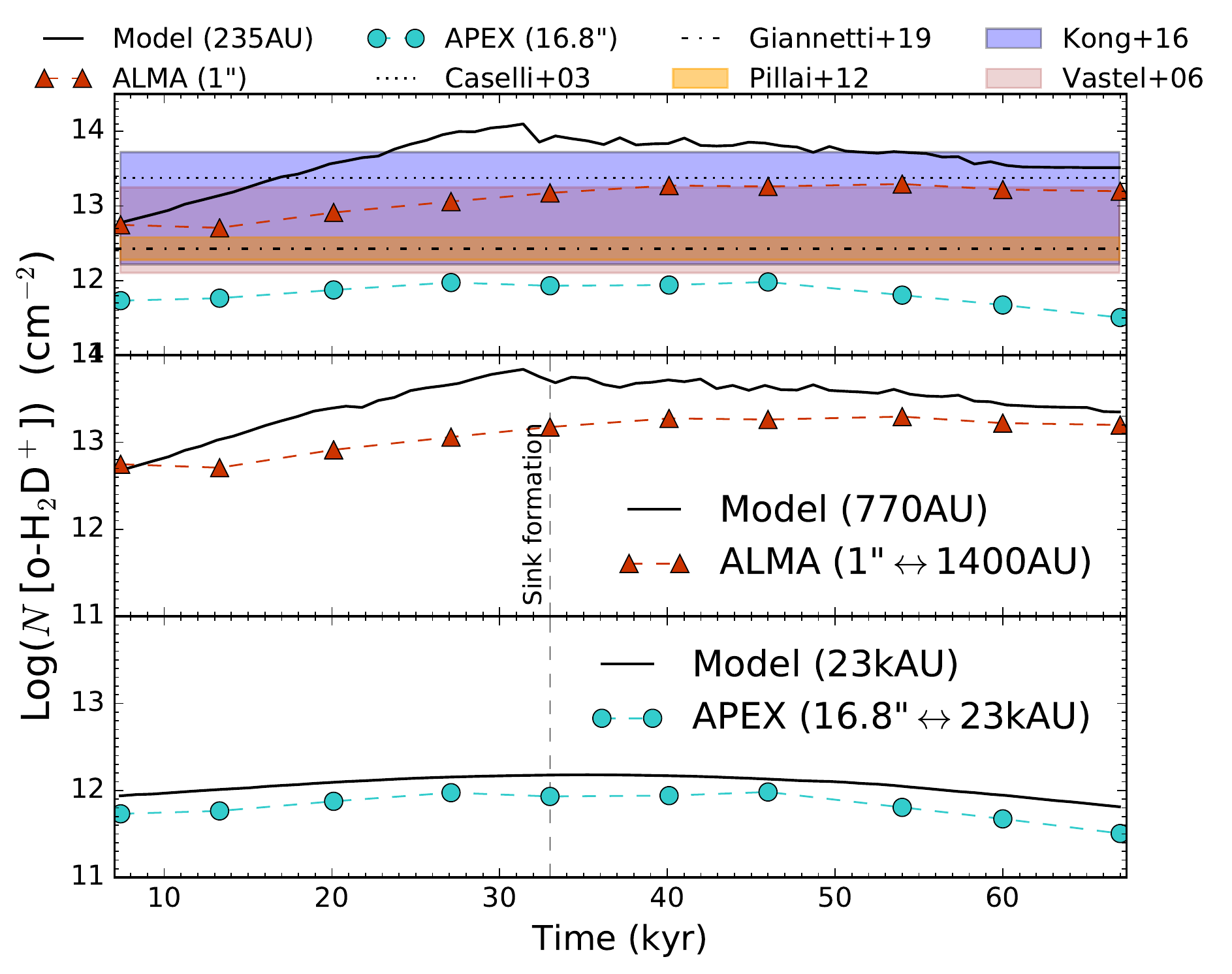}
    \caption{\textit{Top panel:} Column densities derived from the model (solid line) along with interferometric (green pentagons) and single-dish (blue hexagons) observations. Colored areas and horizontal dashed lines represent ranges of values observed onto similar sources; \textit{central and bottom panels:} Comparison of the information loss between ALMA (middle) and APEX (bottom) by averaging the model in the region of each beam.
    }
    \label{fig:column_densities}
\end{figure}

In this work we performed synthetic observations of a collapsing molecular magnetized core at different angular resolutions, attempting to understand the main differences of observing with an interferometer or a single-dish. In the upper panel of Fig.~\ref{fig:column_densities} we show the column densities derived from the model (solid black curve) at 235 AU (highest resolution element in the simulation), from ALMA at 0.55" and from APEX at 16.8", respectively. Column densities at each time were averaged over the highest resolution element. It can be observed that retrieved column densities decrease by 10 (ALMA) and 10$^2$ (APEX) from the model, because, when centered on the peak emission, the inclusion of lower surrounding values is more significant for wider beams, then, decreasing the mean for low resolutions. Therefore, we expect APEX estimations to be lower than those retrieved by ALMA. Colored regions in Fig.~\ref{fig:column_densities} as well as semi-dashed and dotted lines represent observed column densities reported in the literature, obtained via single-dish observations (APEX, JCMT and CSO) toward low-mass sources. 

As a measurement of the information loss, middle and bottom panels of Fig~\ref{fig:column_densities} present the ratio between both resolutions and the model averaged over the area corresponding to each beam. The highest ratios in interferometric observations (middle panel) are up to 10$^{0.5}$. For single-dish observations (bottom panel) the highest ratios are around 10$^{0.3}$. Such ratio decrease for wider beams because the model curve smooths due to the inclusion of surrounding lower values into the average.

\begin{acknowledgement}
JZ and DRGS thank for funding via Fondecyt regular 1161247. DRGS and SB also thank for funding via Conicyt Programa de Astronomia Fondo Quimal 2017 QUIMAL170001 and BASAL Centro de Astrof\'isica y Tecnolog\'ias Afines (CATA) PFB-06/2007. SB thanks for funding through Fondecyt Iniciacion 11170268. SB and JZ also thanks for funding through the DFG priority program “The Physics of the Interstellar Medium” (project BO 4113/1-2). The simulations were performed with resources provided by the \emph{Kultrun Astronomy Hybrid Cluster} at the Department of Astronomy, Universidad de Concepci\'on. We also thank Robi Banerjee, Bastian K\"ortgen, Stefan Reissl and Sebastian Wolf for important contributions to the presented study.
\end{acknowledgement}

\vspace{-0.3cm}

\bibliographystyle{baaa}
\small
\bibliography{biblio}
 
\end{document}